# Custom frame synchronization for easy and rapid deployment

Dimitris Nikolaidis *(dimnikolaidis@mail.ntua.gr)*
*(National Technical University of Athens)*

*Abstract*—We propose a novel and efficient, custom frame synchronization architecture aimed at rapid deployment on any hardware platform. Frame synchronization is the process of discerning valid data frames from an incoming transmission and in this article it is accomplished by attaching distinctive binary overhead sequences on the frame. These sequences act as markers for the frames and enable the capture of their payload. They have certain properties and can be detected by using only simple hardware constructs like XNOR gates and few-bit adders with adequate accuracy. A low-cost commercial FPGA was used for implementation (NEXYS 4 DDR).

*Index Terms—frame synchronization, hardware acceleration, FPGA, low cost, high bit rate*

## I. INTRODUCTION

Frame synchronization is the act of detecting and capturing valid data frames from an incoming transmission and it is one of the most important stages of digital communications. In this article we present a simple frame synchronization hardware architecture which can be implemented on any platform with little effort and resources.

The architecture achieves synchronization by attaching special binary overhead sequences in front of the payload. These sequences are of the form $a_k b_l c_k$ (transmitted from right to left, $c_k$ is first and $a_k$ last) where $a_k$ and $c_k$ are sequences of length $k$ bits and $b_l$ is a sequence of length $l$ bits with $l > 2 \times k$. $b_l$ can be random but $a_k$ and $c_k$ must be the last and first $k$ bits (respectively) of $b_l$ inversed ($a_k$=NOT(last $k$ bits of $b_l$), $c_k$ = NOT (first $k$ bits of $b_l$) ). The architecture only detects $b_l$. $a_k$ and $c_k$ exist to significantly enhance its detection capabilities. The form of the entire frame is $payload_p a_k b_l c_k$, transmitted from right to left. $Payload_p$ is the data bits of the frame (length $p$ bits) and $a_k b_l c_k$ is the extra bits needed for detection by the architecture and can be considered as an atypical header. Detection of the $b_l$ sequence is based on three principles:

1. To calculate the number of same elements in respective positions between two vectors $v_1$, $v_2$ of the same size $s$, we apply the xnor operation on elements in respective positions ($j$ position in both $v_1$ and $v_2$) and we add the products. The result of the summation $sum$ represents how many same elements the two vectors have. When the value of $sum$ is high then the two vectors are similar. When the value of $sum$ is low then the two vectors are not similar. Essentially, it is the opposite of Hamming distance.
2. If we have a much larger vector $v_3$ of size $z$ where $z >> s$ we can use the process in (1) to determine whether $v_1$ exists inside $v_3$. This is achieved by applying (1) in all possible positions $v_1$ can be in inside $v_3$. There are $z-s+1$ possible positions for $v_1$ inside $v_3$ which are $s-1$ down to 0, $s$ down to 1, $s+1$ down to 2…. $z-2$ down to $z-s-1$ and $z-1$ down to $z-s$. This means that (1) is applied $z-s+1$ times and for every iteration $v_1$ is $v_1$ and $v_2$ corresponds to the vector constructed by elements inside $v_3$ of the respective positions (for example elements in positions $s-1$ down to 0 for the first iteration). The process produces $z-s+1$ values $sum_i$. If any of these values equals $s$ or is close to $s$ then this means that $v_1$ or a vector that resembles it was detected inside $v_3$. This process works no matter how large size $z$ of $v_3$ is.
3. For the architecture, $v_3$ is the incoming transmission bitstream and $v_1$ is sequence $b_l$. It detects $b_l$ inside $v_3$ using (2) and captures the payload of the frame.

These 3 principles form the basis for the function of the architecture and henceforth they will be referenced with parenthesis ((1) referring to principle 1). The input of the architecture is the digital stream which results from demodulation of the incoming transmission. Without loss of generality, we assume that the input is given as a parallel bit-vector of size $l$ (same size as sequence $b_l$). In other words, instead of receiving one bit at $f$ frequency the architecture receives $l$ bits at $f_{op} = \frac{f}{l}$ frequency and its total bit rate is $f_{op} \times l$. We also assume that if the parallel input bit vector elements are $l-1$ down to 0 the bit in position 0 was transmitted earlier in the incoming bitstream timeline than bit in position $l-1$. Figure 1 is the visual representation of how the architecture "sees" the incoming datastream.

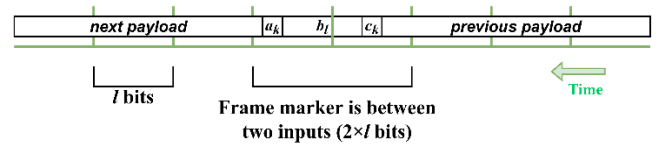

Figure 1: Real time reception of the input. It is separated by blocks of $l$ bits. The frame header sequence $a_k b_l c_k$ ($<2 \times l$) is between two inputs.

Time flows to the left so samples to the right were transmitted earlier. The bitstream is separated into blocks of $l$ bits ($l-1$ down to 0) which is the size of the input. We can clearly see that $b_l$ is always between two inputs ($2 \times l > l + 2 \times k$ bits). The architecture isolates these $2 \times l$ bits each time, correlates them with the predefined $b_l$, finds the start of the payload and captures it. The rare case when the $b_l$ is perfectly aligned with an input sample is also considered.

Using digital correlation as a frame synchronization mechanism is not new. It is in fact the traditional method [1][2] which is still used even today albeit heavily modified [3][4]. Other modern frame synchronization includes joint error correction and frame detection [5][6] (correcting the frame while simultaneously detecting it) and maximum likelihood estimation [7][8] (estimate the most probable position of the frame). Joint error correction and detection together with maximum likelihood produce good results but are complex to implement. The traditional method is simple but is not as accurate. The proposed architecture fills the gap between them by being very simple to implement while providing good detection accuracy.

The paper is separated into 4 sections. The second section is dedicated. The second section contains the description of the architecture along with the description of its function. The third section presents the result of the implementation and testing. The article ends with a conclusion.

## II. ARCHITECTURE

The architecture is comprised of three modules. Window Module, Correlation Module and Payload Capture Module. The Window Module consists of a $2 \times l$ bit register that it uses to isolate the area of the incoming stream in which header $a_k b_l c_k$ is located. It sends the contents of this register to the Correlation Module so that the position of the payload can be calculated. The Correlation Module correlates the predefined stored $b_l$ sequence with the area isolated by the Window Module (in accordance with (2)) and calculates two values $Sum_m$ and $m$. $Sum_m$ is the maximum value of all correlation values in the current window (of $2 \times l$ bits) and $m$ is the position it was detected in. Apart from these two values, the Correlation Module also provides the synchronized input i.e the contents of the Window Module register, for which the values are calculated, to the Payload Capture Module for correct synchronization. Payload Capture Module uses $sum_m$ and $m$ to determine the position of the payload and the synchronized input to capture it. A valid out signal is used to notify the rest of the system of successful capture. The entirety of the receiver architecture can be seen in Figure 2.

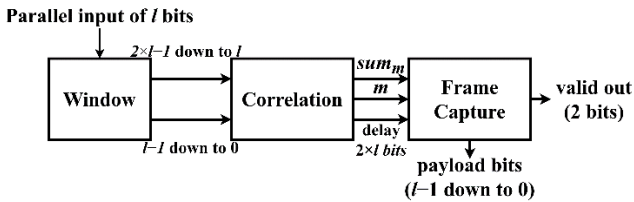

Figure 2: Overview of the system.

### A. Window Module

The Window Module isolates a specific region of the incoming bitstream so that the Correlation Module can determine whether $b_l$ (and as a result the frame) was detected in that specific region. The module consists of a $2 \times l$ bit register (memory elements $2 \times l - 1$ down to 0). The *l-bit* input enters the left half of the register (elements $2 \times l - 1$ down to $l$) from the input pins in clock cycle $t$. In clock cycle $t+1$ contents of the left half are transferred to the right half (elements $l-1$ down to 0) while keeping their order and the new bits enter the left half. Essentially memory elements (D flip flops) that are $l$ positions apart are connected serially. Window module can be seen in Figure 3.

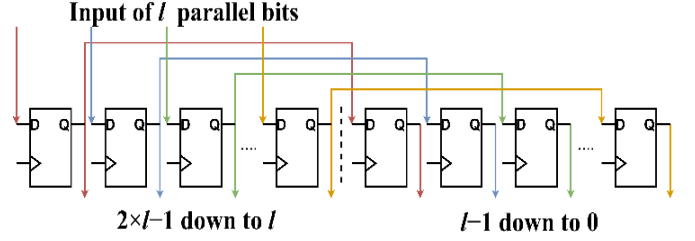

Figure 3: The figure is color coded. Elements connected with the same-colored wire are connected serially.

$2 \times l$ is the minimum number of memory elements needed to capture the $b_l$ sequence when the size of the parallel input is $l$ bits long. We can easily show this by considering the following. If the input vector is $l$ bits long, the starting position of $b_l$ can be any one of these $l$ bits which means that all the possible positions for $b_l$ are: $2 \times l - 2$ down to $l - 1$ ($b_l$ starts from bit $l-1$), $2 \times l - 3$ down to $l - 2$ ($b_l$ starts from bit $l-2$) … $l-1$ down to 0 ($b_l$ starts from 0). Because in every clock cycle elements shift $l$ bits, positions $2 \times l - 1$ down to $l$ is the same as $l-1$ down to 0 (shifted to the right by $l$ bits) and do not need to be checked. Determining which one of these positions is the most probable one is the role of the Correlation Module.

### B. Correlation Module

The Correlation Module is the core of the architecture and is comprised of three main subcircuits. The Parallel Adder Trees, the Selector and the Input Delay Buffer.

The Parallel Adder Trees subcircuit performs the action described in (2). It correlates, in parallel, (digitally as described in (1)) the stored $b_l$ sequence with the vectors in all possible $b_l$ positions. As mentioned, these positions are $2 \times l - 2$ down to $l - 1$, $2 \times l - 3$ down to $l - 2$ … $l - 1$ down to 0. The result of this process is $l$ values $sum_i$, the highest of which is the most probable position for $b_l$. For each position, the correlation value $sum$ is calculated by using xnor gates (for the xnor operation) and an adder tree (to calculate the summation of all xnor outputs). Adder trees are common structures and are extensively used in many hardware designs for calculating the summation of multiple values. In our case, the adder trees have the characteristic form seen in Figure 4.

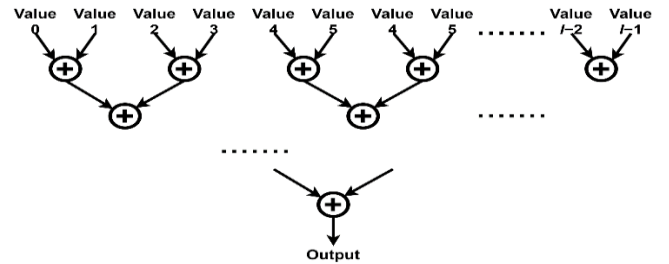

Figure 4: Adder tree.

For the summation of $l$ values, the trees are comprised of

ceil($\log_2 l$) levels. In every level, the values calculated by previous levels are added by two and the output of this process is given as an input to the next level. If the number of values is not a power of two each level is completed by adding zeros to the value that remains. Eventually the summations of all elements are calculated on the last level. By adding registers in-between the adders the structure becomes pipelined. Each adder tree corresponds to one value $sum_i$ so in total $l$ adder trees are used. The latency of the adder tree is ceil($\log_2 l$) clock cycles.

The Selector subcircuit determines ("selects") the highest value $sum_m$ which represents the most probable position of the $b_l$ sequence along with its position indicated by $m$. The selector has a similar form to that of the adder tree however instead of adders the selector tree has comparators which compare two values and output the highest to the next level. There are ceil($\log_2 l$) levels and at the last level the Selector produces the highest value $sum_m$ as the final output. Position value $m$, which essentially corresponds to the position of the first bit of $b_l$, is kept on separate register and advances together with its corresponding value. The (pipelined) Selector has the same latency as the adder tree of ceil($\log_2 l$) clock cycles.

An example of the functionality and circuitry of the Window Module together with the Parallel Adder Trees and the Selector when $l=8$ (length of $b_l$) and $k=3$ (length of $a_k$, $c_k$) can be seen in Figure 5. These lengths, although short, can clearly demonstrate how the frame synchronization process is executed. In the specific example the sequence being searched ($b_l$) is 10001110. For larger $l$ size the circuit depicted is identical. The only difference is a larger window register and more correlators. Ideally $k$ should not be larger than 20% of $l$.

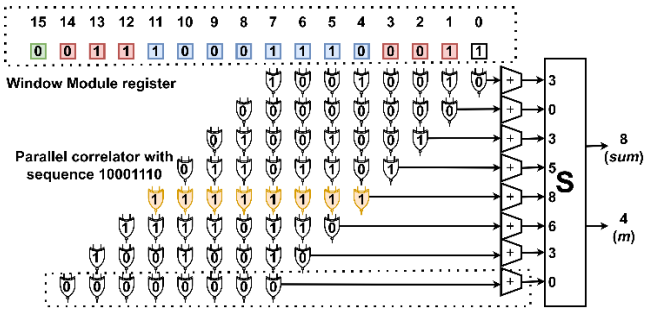

Figure 5: Parallel correlation with window register. The numbered blocks on top represent the window module register. "S" block is the selector circuit. The starting bit of the $b_l$ sequence (10001110) was detected on bit 4 (output $m$). Red squares represent sequences $a_k$ and $c_k$ (with $k=3$) which are the inverted edges of $b_l$ (blue squares). Green square is the start of the payload.

The parallel input of $l=8$ bits first enters the left half (15 down to 8) of Window Module on cycle $t$. On cycle $t+1$ the next input enters the left half, and the previous input migrates to positions 7 down to 0 which brings the start of $b_l$ to position 4 of the overall register (was previously in the 12$^\text{th}$ element). This is the cycle depicted in Figure 5. For subsequent clock cycles the elements inside the window register enter the correlation module. They are correlated with the stored sequence $b_l$ according to (2). Each line of xnor gates together with the adder tree at the end constitutes a correlator. There are 8 correlators. The correlation module produces the maximum value $sum_m$ along with its position $m$ at any given time. Position $m$ represents the starting bit of $b_l$ and it is what Payload Capture Module utilizes to capture the payload from the incoming frame. In Figure 5 $m=4$ which means that the starting bit of the payload is $l+k+m=8+3+4=15$(green square). The latency of the correlation module is ceil($\log_2 l$)+ceil($\log_2 l$)=2×ceil($\log_2 l$) (Selector and adder tree).

Since the Correlation Module has a latency of 2×ceil($\log_2 l$) the Payload Capture Module (next section) cannot use the Window Module register for payload capture as the inputs are not synchronized with the outputs. To capture the payload data effectively the Payload Capture Module needs to be able to "see" the window register with a delay synchronized to the corresponding outputs of the Correlation Module. To do this the Correlation Module is equipped with a delay buffer. The buffer causes a delay of 2×ceil($\log_2 l$) clock cycles to the input and synchronizes it with the corresponding values $sum_m$ and $m$. The delayed input together with the two values are given to the Payload Capture Module to enable payload capture.

### C. Payload Capture Module

The Payload Capture Module utilizes the two outputs of the correlation module $sum_m$ and $m$ together with the delayed input to correctly deduce the position of the payload and capture the data. As mentioned in (1) when the value of the digital correlation is high then the two binary vectors are close in terms of likeness. Likewise, if the value of $sum_m$ (which is the highest of all positions) is above a certain threshold then this means that the correlator has located a vector at position $m$ that is very close to $b_l$. If the threshold is appropriately set the Payload Capture Module can not only detect the payload but also capture it regardless of frame size. For this reason, selecting the proper value for the threshold is instrumental for the correct operation of the receiver.

To determine the starting position of $b_l$ and the payload, the module uses a simple control unit which constantly checks output $sum_m$ of the Correlation Module. When the output surpasses the threshold at clock cycle $t$ the control unit is activated. It assumes that the payload is found at $l+k+m_t$ and begins its capture. In the next cycle $t+1$, it compares the new value of $sum_{m_{t+1}}$ with the previous $sum_{m_t}$. If $sum_{m_{t+1}} < sum_{m_t}$ then the first bit of $b_l$ was at position $m_t$ and no changes are necessary. If the opposite is true, the first bit of $b_l$ is at position $m_{t+1}$ and the frame capture process must restart to capture the correct bits. The position of the payload in relation to the frame header can be seen in Figure 6.

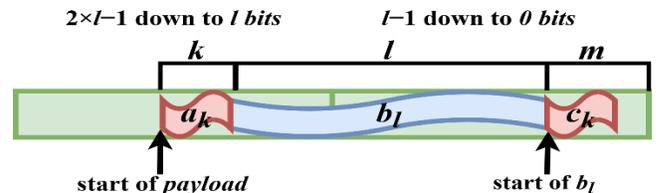

Figure 6: Position of payload in relation to the frame header $a_k b_l c_k$.

There are multiple ways in which capture can be achieved but in this article we propose the simplest. If we examine Figure 6, we conclude that because $b_l$ is $l$ bits long $l+k+m_t$ (or $l+k+m_{t+1}$) will either be outside the register or inside the left half ($2 \times l-1$ down to $l$). If it is inside the register the capture process begins immediately. If it is not, the control unit waits one cycle to start the process. Assuming that the size of the payload is a multiple of $l(p=n \times l)$ the payload can be captured in $n+1$ cycles in the following manner. When the first bit of the payload enters the left half of the register the unit captures the last $l-(k+m_t)$ (or $m_{t+1}$) bits of the register (left half) which are the first bits of the payload. For the next $n-1$ cycles all bits in the left half are captured. In the final cycle the first $k+m_t$ (or $m_{t+1}$) bits of the left half are captured. If we add all the bits together we have $l-(k+m_t)+(n-1) \times l+k+m_t=n \times l$ bits which is the entire payload. The payload is captured at the same rate as it enters the circuit and the two-bit signal valid out is used to indicate that the capture process is underway. The biggest advantage of using this method to capture the frames is that no matter how large $n$ is (the size of the frame) the size of the circuit does not change, only the cycles needed to capture it. The user can disable the correlation module while capturing is in progress to reduce power consumption.

### III. IMPLEMENTATION, TESTING AND ASSESSMENT

The receiver was implemented on the low cost NEXYS 4 DDR FPGA board. The implementation was for $l=123$, $k=23$, $th=89$(threshold). A maximum operating frequency of 125Mhz was achieved. The bit rate was obtained by multiplying the parallel input with the operating frequency and we have 123bits×125MHz=15.37Gbps. The needed hardware resources, utilization, power consumption and bit rate are all presented in Table 1.

Table 1: Implementation Results.

| $b_l$ | LUTs (%) | FFs (%) | P(W) | Bit Rate (Gbps) |
|---|---|---|---|---|
| $l=123$ | 20027(32) | 23608(18) | 0.842 | 15.375 |

VIVADO and MATLAB were used for simulations. Input bitstreams were prepared in MATLAB and inserted in the VHDL testbench as txt files. The modulation scheme of choice was OFDM with 128 16QAM subcarriers and range of SNR was [-8, 2]. Inputs were prepared for each SNR integer value and version separately. They consisted of 21368 frames of the form $payload_j a_k b_l c_k$ (as described in the section II). The bitstreams were created, modulated with OFDM, driven through the AWGN channel (MATLAB simulation). After this process, the bitstreams were demodulated, recorded in txt files and given as input to the implemented circuit with a VHDL testbench. The implemented design outputs the payload of each frame. The captured payloads were of size 12300(multiple of 123) and were checked by MATLAB. The frame synchronization error rate (FSER) can be seen in Figure 7.

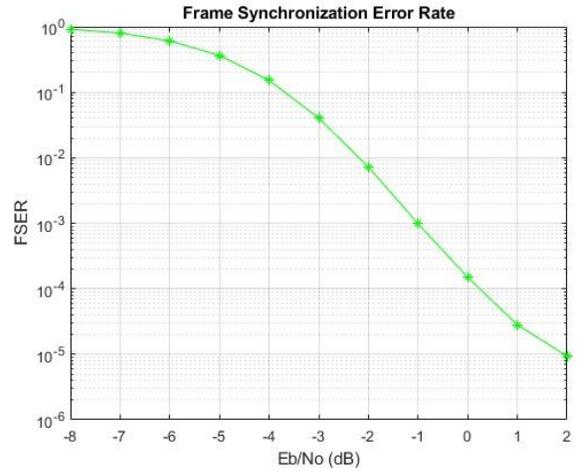

Figure 7: Frame synchronization Error Rate of the architecture.

We see that the circuit achieves accuracy which rivals other state of the art hardware (2~3 dB worse) [9] while being very simple and easy to implement. Moreover, the fact that the frame marker($b_l$) used to detect the frame itself can be random in nature means that the design can be used as physical layer security mechanism to ensure transmission security [10][11]. Also, it can be implemented on an ASIC platform to achieve even higher line rate (by having faster clock) with even less hardware and power consumption [12] (on average ×4-line rate and ÷35 hardware).

### IV. CONCLUSION

In this paper we have presented a simple yet accurate novel frame synchronization architecture. The architecture can be implemented on very low-cost platforms (NEXYS 4 DDR FPGA board), consumes little in terms of hardware and power while providing bit rates of up to 15.375 Gbps with adequate accuracy. Its minimalist nature allows for rapid implementation and its flexibility enables it to be modified easily. Future work involves modified versions aimed at maximizing specific traits such as accuracy, bit rate and security.